# Theoretical Modelling of High-Resolution X-Ray Absorption Spectra at Uranium $M_4$ Edge

Jindřich Kolorenč[1] and Kristina O. Kvashnina[2,3]

[1]*Institute of Physics, Czech Academy of Sciences, Na Slovance 2, 182 21 Praha, Czech Republic*

[2]*Institute of Resource Ecology, Helmholtz-Zentrum Dresden-Rossendorf, 01314 Dresden, Germany*

[3]*Rossendorf Beamline at The European Synchrotron (ESRF), CS40220, 38043 Grenoble Cedex 9, France*

**ABSTRACT**

*We investigate the origin of satellite features that appear in the high-resolution x-ray absorption spectra measured at the uranium $M_4$ edge in compounds where the uranium atoms are in the $U^{6+}$ oxidation state. We employ a material-specific Anderson impurity model derived from the electronic structure obtained by the density-functional theory.*

## INTRODUCTION

The resolution of the x-ray absorption spectroscopy is fundamentally limited by the life-time broadening of the core hole created in the absorption event. For the actinide M edges (3d → 5f transitions), this broadening is large due to a short life time of the deep 3d hole (∼3.5 eV [1]). However, when the absorption is detected by monitoring a particular core-to-core process of filling the created hole (the 4f → 3d transition in the case of the actinide M edge), the broadening is reduced and it is determined by a considerably longer life time of the shallower 4f hole (∼0.3 eV [2]) [3]. This method is referred to as the high-energy-resolution fluorescence-detected x-ray absorption spectroscopy (HERFD-XAS) in the literature [4]. The spectra recorded in this way do not exactly coincide with the conventional x-ray absorption: XAS involves only one photon and can be calculated from the Fermi golden rule, whereas HERFD-XAS involves two photons and thus should be modelled by the Kramers–Heisenberg formula. Nevertheless, the Kramers–Heisenberg formula simplifies to an expression very similar to the conventional XAS formula in certain cases. One set of such simplifying assumptions was recently discussed in the context of the resonant x-ray emission spectroscopy (RXES) at the L edge of lanthanides [5]: If (i) the shape of the core orbitals can be neglected so that the core-valence Coulomb interaction is fully determined by just one Slater integral, (ii) if this core-valence Slater integral is the same in the intermediate state (3d hole) and in the final state (4f hole), and (iii) if the detector registers all polarizations of the emitted photon equally, then the HERFD-XAS essentially coincides with the conventional XAS, only the life-time broadening is described by a function that slightly differs from the Lorentzian. In the case of the L-edge RXES of praseodymium, this simplified model achieves a very good agreement with experiment [5]. The calculations reported in the present paper are performed under the same assumptions.

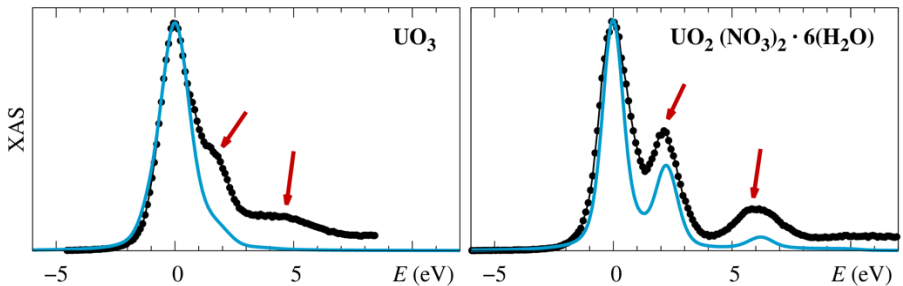

**Figure 1.** The experimental HERFD spectra at the uranium $M_4$ edge of $UO_3$ and uranyl nitrate (black dots) are compared to $j = 5/2$ component of the 5f-electron density of states computed in the local-density approximation (blue line). The LDA density of states is broadened with a Voight convolution of a Gaussian representing the instrument resolution (FWHM of 1.0 eV) and a Lorentzian representing the life-time effects (FWHM of 0.3 eV). The broadening parameters are the same for all theoretical spectra shown in this paper. The satellite features discussed in the text are indicated with arrows.

We investigate absorption at the uranium $M_4$ edge ($3d_{3/2} \rightarrow 5f_{5/2}$) in compounds where the uranium atoms are in the $U^{6+}$ oxidation state. The compounds in question are $\alpha$-$UO_3$ and uranyl nitrate $UO_2(NO_3)_2 \cdot 6(H_2O)$. The experimental spectra are shown in Figure 1. Two extra features are found above the main absorption line in both compounds: two shoulders in $UO_3$ and two well defined satellites in uranyl nitrate. Very similar spectrum to uranyl nitrate is measured also for other compounds containing the $(UO_2)^{2+}$ molecular ion [6,7]. It suggests that this spectral shape is essentially determined by this uranyl-type of ion.

## COMPUTATIONAL METHOD

The absorption spectra are computed by exact diagonalization of an Anderson impurity model that represents the uranium 5f shell and its environment [4]. The parameters of this impurity model are derived from first-principles electronic structure [8,9]. In particular, we employ methods previously implemented for LDA+DMFT calculations of actinide dioxides, see Ref. 10 for technical details of the construction of the model as well as of its exact diagonalization. First, the band structure is computed in the local-density approximation (LDA) with the aid of the WIEN2K code [11] and then the relevant bands are represented by a tight-binding hamiltonian in the basis of the corresponding Wannier functions [12,13]. The local electronic structure around one shell of uranium 5f Wannier functions is subsequently mapped onto a non-interacting impurity model that consists of 14 spinorbitals of the 5f shell and another 14 spinorbitals representing the environment dominated by the oxygen 2p states [10]. The construction of the impurity model suitable for calculation of core-level spectra is finished by introducing a spherically symmetric Coulomb vertex acting among the 5f orbitals (this vertex is parametrized by four Slater integrals) and by adding the appropriate core orbitals and the corresponding core-valence Coulomb interaction parametrized by a single Slater integral $U_{cv} = F_0(4f,5f) = F_0(3d,5f)$. The Slater integrals are chosen empirically since we are not able to treat the screening effects from first principles at present. The valence Coulomb parameters are set the same as in the earlier study of actinide dioxides $UO_2$, $NpO_2$ and $PuO_2$ [10], that is, $U = F_0(5f,5f) = 6.5$ eV, $F_2(5f,5f) = 8.1$ eV, $F_4(5f,5f) = 5.4$ eV, and $F_6(5f,5f) = 4.0$ eV. In Ref. 10, it was also shown that the LDA+DMFT impurity model accurately reproduces the x-ray photoemission from the 4f level in these oxides if the core-valence Slater integral $F_0(4f,5f)$ is set to 6 eV. The integral for the 3d level $F_0(3d,5f)$ should be slightly larger and since our assumptions listed above require $F_0(4f,5f)$ and $F_0(3d,5f)$ to be equal, we settle on a compromise value $F_0(4f,5f) = F_0(3d,5f) = 6.5$ eV.

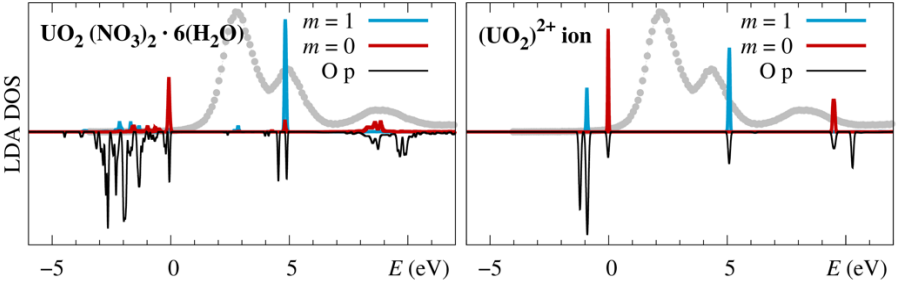

**Figure 2.** The origin of the satellite features in uranyl nitrate is illustrated using similarity to a lone $(UO_2)^{2+}$ ion. The LDA density of states (with the spin-orbit coupling switched off) is projected onto uranium 5f functions with $|m=1\rangle$ and $|m=0\rangle$ character, and onto 2p functions of the two nearest-neighbor oxygen atoms.

Majority of the calculations presented in this paper are performed for the experimental crystal structures. The $\alpha$ phase of $UO_3$ has the space group $P\bar{3}m1$, the uranium atoms are located in 1a positions and the oxygen atoms occupy 2d ($z = 0.17$) and 1b positions. The uranyl nitrate crystallizes in the low-symmetry $Cmc2_1$ structure. It has two formula units (58 atoms) in the primitive cell, too many to list them all here. Only the results shown in Figure 1 and in Figure 2 employ this full structure of uranyl nitrate. The impurity model used to calculate the spectrum in Figure 3 was constructed for a simplified structure: the water molecules were removed and the $UO_2(NO_3)_2$ clusters were placed on an fcc lattice. This higher-symmetry artificial structure accurately represents the environment around the uranium atoms using only 11 atoms in the primitive cell. The LDA band gap of the new structure is noticeably smaller than the gap of the full structure but the unoccupied 5f densities of states of the two structures are almost indistinguishable.

The LDA electronic structure found with the WIEN2K code incorporates scalar-relativistic effects as well as the spin-orbit coupling. The calculations of $\alpha$-$UO_3$ were performed with the following parameters: the radii of the muffin-tin spheres were $R_{MT}(U) = 2.20\ a_B$, and $R_{MT}(O) = 1.70\ a_B$, and the basis-set cutoff $K_{max}$ was defined with $R_{MT}(O) \times K_{max} = 7.00$. The calculations of the uranyl nitrate in the full structure employed muffin-tin spheres with radii $R_{MT}(U) = 2.12\ a_B$, $R_{MT}(O) = 1.06\ a_B$, $R_{MT}(N) = 1.11\ a_B$, and $R_{MT}(H) = 0.57\ a_B$, and the basis-set cutoff was given by $R_{MT}(H) \times K_{max} = 2.57$. Finally, for the simplified structure we used $R_{MT}(U) = 2.12\ a_B$, $R_{MT}(O) = 1.16\ a_B$, $R_{MT}(N) = 1.11\ a_B$, and $R_{MT}(N) \times K_{max} = 5.00$. The relatively small muffin-tin spheres around uranium and oxygen atoms in the uranyl nitrate structures are enforced by short U–O and O–H bonds.

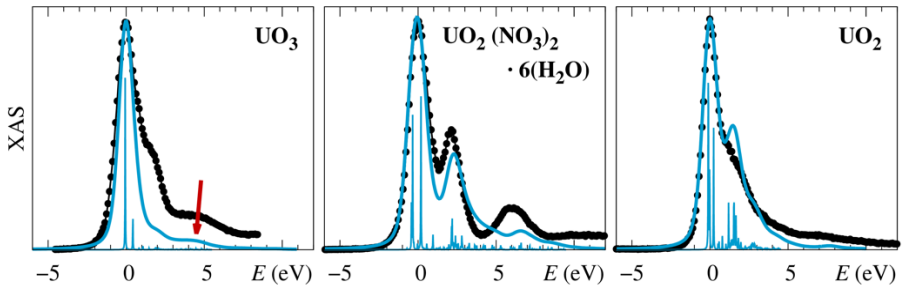

**Figure 3.** The x-ray absorption spectra calculated in the material-specific impurity model. Apart from the two $U^{6+}$ compounds, the uranium dioxide, where uranium is nominally in the oxidation state $U^{4+}$, was added for comparison.

## RESULTS AND DISCUSSION

### Local-density approximation

The simplest approximation to the $M_4$ absorption spectrum is the density of unoccupied $5f_{5/2}$ states that can be computed using the density-functional theory (DFT). The densities of states obtained in the local-density approximation are shown in Figure 1 alongside the experimental HERFD-XAS spectra (the details of the measurements are described elsewhere [6]). The theory reproduces the spectrum of uranyl nitrate fairly well, both satellites are present, albeit their intensity is underestimated. When applied to the plutonyl ion, $(PuO_2)^{2+}$, the DFT achieves a similarly good agreement with experimental data (M-edge HERFD-XAS of $(PuO_2)^{2+}$ displays a single high-energy satellite, intensity of which is again underestimated by the DFT) [14]. The LDA is not quite as successful in $UO_3$ where only a featureless main line is predicted. There is a faint hint of a shoulder at about 2 eV above the main line that comes from the topmost 5f band that, unlike the other 5f bands, shows some dispersion. It is unlikely, however, that this would be the true origin of the experimentally observed shoulder in $UO_3$.

The satellites in uranyl nitrate deserve a closer inspection. In Figure 2 we plot LDA densities of states computed with the spin-orbit coupling switched off to make things simpler and more transparent. The uranyl nitrate is compared to a lone $(UO_2)^{2+}$ molecular ion to demonstrate that the satellite features indeed originate in this ion. The environment of the uranium atom is highly anisotropic in this ion and since the molecule is linear, it is useful to project the 5f density of states onto the angular-momentum eigenstates $|l = 3, m\rangle$ with the quantization axis pointed along the O–U–O bonds. The main line consists of $|m = \pm 2\rangle$ and $|m = \pm 3\rangle$ states that point away from the oxygen atoms (not explicitly shown), and the satellites have $|m = \pm 1\rangle$ and $|m = 0\rangle$ character. Each of the $|m = \pm 1\rangle$ and $|m = 0\rangle$ orbitals hybridizes with the oxygen 2p states and forms a bonding combination that is occupied and an antibonding combination that is empty and shows up as a satellite in the absorption spectra. The $|m = \pm 1\rangle$ states ($5f_{xz^2}$, $5f_{yz^2}$, $2p_x$, $2p_y$) make up π molecular orbitals and the $|m = 0\rangle$ states ($5f_{z^3}$, $2p_z$) form σ molecular orbitals. We thus arrive at the same identification of the satellite features as the recent literature [7,14]. The distance between the bonding and antibonding combinations (about 6 eV for $|m = \pm 1\rangle$ and as much as 9 eV for $|m = 0\rangle$) measures the hybridization strength between the corresponding 5f and 2p orbitals.

**Anderson impurity model**

The DFT approximation to the absorption spectra shown in Figure 1 ignores the final-state effects induced by the presence of the core hole. These effects are not necessarily small: the impurity-model calculation of the absorption spectrum of uranyl acetylacetonate presented in [15] indicates that a core-hole-induced shake-up tail possibly extends all the way up to the $m = 0$ satellite (antibonding σ orbital) located 6 eV above the main line. The impurity model used in that study was spherically symmetric and thus it could not correctly reproduce the satellites discussed in the preceding paragraph. Here we present improved impurity-model calculations that combine the core-hole effects with a realistic anisotropic hybridization and crystal field. The results are shown in Figure 3.

In $UO_3$, the impurity model improves the agreement between the theory and experiment compared to Figure 1. The calculations suggest that the higher-energy shoulder (indicated with an arrow in Figure 3) is a result of the shake-up processes. The shoulder at 2 eV above the main line is not reproduced and it remains a puzzle. In uranyl nitrate, the core-hole effects slightly widen the main line and the shake-up tail overlaps with the satellites that still have their intensity underestimated similarly to the LDA calculation.

In addition to the two $U^{6+}$ compounds, the absorption spectrum of $UO_2$, a nominally $U^{4+}$ compound, is shown in Figure 3 for comparison. This calculation uses the impurity model constructed as a part of the LDA+DMFT study reported in [10]. In this case, the shoulder at the high-energy side of the absorption line is identified with a valence $5f^2$ multiplet. The feature remains nearly intact when the impurity model is reduced to a spherically symmetric atomic model by removing the hybridization with the ligand states and the crystal field. The $UO_2$ calculation is included for illustration of the general applicability and universality of the presented computational method.

**CONCLUSIONS**

The satellites appearing in the $M_4$-edge x-ray absorption spectrum of uranyl nitrate are identified with antibonding σ and π molecular orbitals consisting of uranium 5f and oxygen 2p states that are located at the U–O bonds in the $(UO_2)^{2+}$ molecular ion. This finding rationalizes why analogous satellites are observed also in other similar compounds like uranyl acetylacetonate and torbentite, and it provides a quantitative microscopic backing to the mechanism that was suggested on an empirical basis earlier [16,17]. The presented results illustrate that the material-specific impurity model derived from the LDA or LDA+DMFT band structure is a quite versatile tool: it reproduces valence-band many-body features ($UO_2$), single-particle features due to anisotropic environment (uranyl), as well as core-hole-induced shake-up features. However, our current implementation offers only a partial understanding of the absorption spectra of $UO_3$ (the 2 eV shoulder is not reproduced) and an improved theoretical investigation will need to be performed in the future.

**ACKNOWLEDGMENTS**

We acknowledge support by the Czech Science Foundation under the grant number 18-02344S (JK), and by a starting grant from the European Research Council (ERC), number 759696 (KOK). Access to computing facilities owned by parties and projects contributing to the National Grid Infrastructure MetaCentrum, provided under the program Cesnet LM2015042, is appreciated.